\documentstyle[11pt,epsf]{article}

\catcode`@=11
\def\seceqaa{\@addtoreset{equation}{section}
\def\theequation{A\arabic{equation}}}
\def\seceqbb{\@addtoreset{equation}{section}
\def\theequation{B\arabic{equation}}}
\def\seceqcc{\@addtoreset{equation}{section}
\def\theequation{C\arabic{equation}}}
\catcode`@=12

\newcommand{\pdot}{{\displaystyle{\raisebox{-1.5ex}[0.25ex]{$\cdot$}
     \atop\raisebox{0.6ex}[0.25ex]{$\scriptstyle (p)$}}}}
\newcommand{\prdot}{{\displaystyle{\raisebox{-1.5ex}[0.25ex]{$\cdot$}
     \atop\raisebox{0.6ex}[0.25ex]{$\scriptstyle (p^{\prime})$}}}}
\newcommand{\ppdot}{{\displaystyle{\raisebox{-1.5ex}[0.25ex]{$\cdot$}
     \atop\raisebox{0.6ex}[0.25ex]{$\scriptstyle (p,p^{\prime})$}}}}
\newcommand{\kuroten}{{\displaystyle{\raisebox{-1.5ex}{$\odot$}
     \atop\raisebox{0.3ex}[1.25ex]{$\scriptstyle(p,p^{\prime})$}}}}
\newcommand{\peke}{{\displaystyle{\raisebox{-1.5ex}{$\times$}
      \atop\raisebox{0.3ex}[1.25ex]{$\scriptstyle (p,p^{\prime})$}}}}
\begin{document}
\input{psfig}
\begin{center}
\hfill IP/BBSR/2001-16\\
{\Large \bf 
Noncommutative $N=2$
$p-p^\prime$ System 
}
\vskip 0.1in
{Aalok Misra\\ 
Institute of physics,\\
Bhubaneswar 751 005, India\\email: aalok@iopb.res.in}
\vskip 0.5 true in
\end{center}
\begin{abstract}
We analyse several open and  mixed sector tree-level amplitudes 
in $N=2$  $p-p^\prime$ systems with a constant magnetic $B$ turned on. 
The 3-point function vanishes on-shell. The 4-point 
function, in the Seiberg-Witten (SW)
low energy limit\cite{SW}, is local, {\it indicating the possible
topological nature of the theory (in the SW low energy limit)} and
the {\it possible relation between noncommutative $N=2$ $p-p^\prime$ system
in two complex dimensions and in the SW limit, and (non)commutative $N=2$ 
$p^\prime-p^\prime$ system in two real dimensions.}
We discuss three extreme noncommutativity limits (after having taken the
Seiberg-Witten low energy limit) of the mixed 3-point function,
and get two kinds of commutative non-associative generalized star
products.  We make some speculative remarks related 
to reproducing the above four-point tree level amplitude  in the
open sector, from a field theory. 
\end{abstract}

\section{Introduction}

$N=2$ strings have been studied in the past for a variety of reasons.
They are extremely useful from the point of view of studying self-dual
gravity and Yang-Mills theories \cite{OV,Mar}. Also, they are thought to be
intimately connected to M(atrix) and F theories
\cite{Mart,Siegel,Ketov}. Noncommutative $N=2$ strings were first 
studied in \cite{JHEP1} which showed several interesting features -
(a) appearance of Moyal star product in open-string amplitudes and
hence the topological nature of the purely open (and closed) sector(s),
(b) construction of abelian noncommutative effective field 
theory in the purely open sector (equivalently, 
abelian noncommutative self-dual
Yang-Mills in flat space), unlike its commutative $N=2$ counterpart,
(c) appearance of generalized star product in the mixed 3-point
function,
and (d) vanishing of the mixed 4-point function $A_{oooc}$ in the
extreme noncommutativity limit. 

There are the following three motivating reasons for studying 
noncommutative $N=2$ $p-p^\prime$ system.  (i) The work of \cite{JHEP1} 
had to do with noncommutative $N=2\ p^\prime-p^\prime$ system, 
i.e., the open strings starting and ending on the same brane (or
branes of the same dimensionality with identical boundary conditions
at the two ends of the open string)
in the presence of external magnetic field. It 
is hence natural to extend this to the case where 
the two ends of the open string end on branes of different 
dimensionality (in the presence of a magnetic background), 
i.e., noncommutative $N=2 \ p-p^\prime(>p)$ system. As the boundary 
conditions are different at the two ends of the open string, one 
would expect that the theory is not topological due to the shift in the vacuum 
energy. In this work, we discuss, among other things, {\it whether it 
is possible, in any limit, to get a topological noncommutative $N=2$ theory, at 
least in the purely open sector} 
\footnote{There is no difference between noncommutative $N=2$ $p-p^\prime$ and 
$p^\prime-p^\prime$ theories in the purely closed sector. Further the 
noncommutative $N=2$ $p^\prime-p^\prime$ theory is the same as the
commutative $N=2$ $p^\prime-p^\prime$ theory 
in the closed sector, which is known to be topological \cite{OV}. Hence,
the closed sector of noncommutative $N=2$ $p-p^\prime$ system is topological.}. 
(ii) Secondly, the mixed-sector of noncommutative $N=2$ $p^\prime-p^\prime$ 
system, involved a generalized star product in the mixed 3-point function and 
hence the field theory that would reproduce the string amplitude. It is hence 
natural to ask whether generalized star product(s) of the same or different
type(s) appear in the mixed sector of noncommutative $N=2\ p-p^\prime$ system. 
(iii) Finally, the field theory that reproduced the string amplitudes of
noncommutative $N=2$ $p^\prime-p^\prime$ system
in \cite{JHEP1} consisted of the open-string metric and the Moyal star
product in the purely open sector, and a generalized star product and
two linear combinations of the open-string metric and the noncommutativity
parameter in the mixed-string sector. It is of importance 
to construct the field theory that would reproduce 
the string amplitudes of noncommutative $N=2$ $p-p^\prime$ system.  

We study amplitudes involving either two $p-p^\prime$ 
open strings and one or two $p^\prime-p^\prime$ open strings,
or two $p-p^\prime$ open strings and one closed string, or two 
$p^\prime-p^\prime$ open strings and one closed string.
We summarize our results vis-a-vis the abovementioned three motivating reasons. 
(i) Interestingly, we find that  in the purely
open sector of $N=2\ p-p^\prime$ system, the 3-point function vanishes
on-shell and the 4-point function, in the Seiberg-Witten (SW) low energy limit,
is local. {\it This strongly suggests that the purely open sector of 
noncommutative $N=2$ $p-p^\prime$ system, is topological in the SW low energy 
limit}. (ii) The mixed sector is more non-trivial than in \cite{JHEP1}.
For finite noncommutativity, we get contact-term delta function divergences.
For infinite noncommutativity (taking the infinite noncommutativity limit
in a specific manner), after having taken the SW low energy limit,
we show that the abovementioned divergence can be disregarded, and
depending on whether one takes infinite noncommutativity along the common
and/or the uncommon directions, one gets commutative non-associative 
generalized star products of two kinds in the mixed 3-point 
function. Also, in this limit, one gets 
an infinite series of local interactions for the mixed 3-point function.
(iii) We show that it is not possible to construct a field theory
that would reproduce the local 4-point function of the purely open sector,
using the 3-point functions evaluated in this paper, in the purely open
and mixed sectors. We speculate on possible resolution of this problem.

The paper is organised as follows. In section 2, we evaluate 3-point
and 4-point string amplitudes in the purely open sector of the types 
mentioned above, by reading off  results from the corresponding
expressions given for  $N=1$ $p-p^\prime$ systems 
in the presence of magnetic $B$ in \cite{Itoyama}, after a suitable 
identification. We show that the 3-point function vanishes on-shell,
and the 4-point function, in the SW low energy limit, is local.
In section 3, we evaluate two mixed 3-point functions
of the types mentioned above, for finite and infinite noncommutativity,
in the SW low energy limit. We explicitly show the appearance 
of $(\delta(0))^{n=1,2}$-type divergences 
for finite noncommutativity, which can be taken care of
by taking infinite noncommutativity limit suitably after having taken
the SW low energy limit. We show the appearance of generalized star products
of two kinds in this limit. Section 4 has a summary of and a discussion 
on results obtained in this work. 

\section{Purely open sector}

In this section, we discuss 3- and 4-point string amplitudes involving
vertex operators for two $p-p^\prime$ open strings and one or two
$p^\prime-p^\prime$ open strings, the latter in the SW low energy limit.

In the purely open sector, one can read off results
from \cite{Itoyama} after suitable identification of the polarization
vector that figures in the vector vertex operator of \cite{Itoyama}.
This does not imply that a four-dimensional $N=2$ theory can be mapped
to a ten-dimensional $N=1$ theory. What is implied and hence what gets
used in the calculations below, is that the open-string $N=2$ vertex operators,
and hence the open-string $N=2$ open amplitudes,  can 
be obtained from open-string $N=1$ vertex operators and hence
open-string $N=1$ amplitudes,
after the abovementioned identification; the closed string 
vertex operators that are constructed in this work, 
were not considered in \cite{Itoyama}.

The closed-(\cite{OV}) and open-string (\cite{Mar})
vertex operators, in the notations of \cite{Itoyama} are given by:
\begin{eqnarray}
\label{eq:8}
& & V_c^{\rm int}\sim
\biggl(ik\cdot\partial{\bar x}-i{\bar k}\partial x
-\alpha^\prime k\cdot{\bar\psi}_R{\bar k}\cdot\psi_R\biggr)
\biggl(ik\cdot{\bar\partial}{\bar x}-i{\bar k}{\bar\partial} x
-\alpha^\prime k\cdot{\bar\psi}_L{\bar k}\cdot\psi_L\biggr)
e^{i(k\cdot {\bar x}+{\bar k}\cdot x)}
,\nonumber\\
& & V_o^{\rm int}\sim\biggl(ik\cdot\partial_\tau{\bar x}
-i{\bar  k}\cdot{\partial}_\tau x
-\alpha^\prime k\cdot({\bar\psi}_L+{\bar\psi}_R){\bar k}\cdot
(\psi_L+\psi_R)\biggr)
e^{i(k\cdot{\bar x}+{\bar k}\cdot x)},
\end{eqnarray}
In $N=1$ notations, 
\begin{equation}
\label{eq:vertopequiv1}
V_o^{\rm int} = \int d\eta e^{\biggl(i\sqrt{{\alpha^\prime\over2}}
(k\cdot{\bar X} + i{\bar k}\cdot X)
-\eta(D_L+D_R)(k\cdot X - {\bar k}\cdot X)\biggr)}|_{\theta_L=\theta_R},
\end{equation}
which can be identified with
\begin{equation}
\label{eq:vertopequiv2}
\int d\eta e^{i\biggl(\sqrt{{\alpha^\prime\over2}}
(k\cdot{\bar X} + {\bar k}\cdot X)
  +\eta(D_L+D_R)(\zeta\cdot{\bar X}
+{\bar\zeta}\cdot X)\biggr)}|_{\theta_L=\theta_R}
\end{equation}
of \cite{Itoyama}, 
$X$ being a chiral superfield and $\eta$ 
a Grassmanian parameter and $\zeta$ being the
polarization
vector, by setting $\zeta\equiv i k$. In the above, 
$D_L={\partial\over\partial\theta_L}+\theta_L\partial$ and
$D_R={\partial\over\partial\theta_R}+\theta_R{\bar\partial}$.

We briefly outline the main idea of \cite{Itoyama} when one considers
evaluation of amplitudes for $p-p^\prime$ systems in the presence 
of nonzero $B$. As there is no space-time supersymmetry and no
Ramond-Ramond fields in $N=2$ theory, the $p(p^\prime)$ branes
are branes defined in the sense of open-string boundary conditions:
\begin{eqnarray}
\label{eq:bcs}
& & \partial_\sigma X^{1,{\bar 1}} + (B\partial_\tau X)^{1,{\bar 1}}=0,
{\rm \ at}\ \sigma=0,
\nonumber\\
& & \partial_\tau X^{2,{\bar 2}}=0, {\rm\ at}\ \sigma=0,\nonumber\\
& & \partial_\sigma X^{1,{\bar 1},2,{\bar 2}}
+(B\partial_\tau X)^{1,{\bar 1},2,{\bar 2}}=0,{\rm\ at}\ \sigma=\pi,
\end{eqnarray}
where we have complexified the space-time coordinates.
The $p$-brane is a brane with (2,0) signature on its world volume
and the $p^\prime$-brane is a brane with (2,2) signature on its world
volume. As there is no tachyon in the $p-p$ or $p^\prime-p^\prime$
open strings, and as shown below, there are no tachyons in the
$p-p^\prime$ open strings, hence, these nonsupersymmetric branes are
stable. As the boundary conditions at the ends of the $p-p^\prime$
open strings are different, this implies that the vacuum energy gets
shifted relative to the (non)commutative $p^\prime-p^\prime$ 
theory. One thus has fields in
addition to the massless scalar of the $p^\prime-p^\prime$ 
(non)commutative $N=2$ strings. As explained in \cite{Itoyama},
one has to introduce ``shift'' $\sigma^\pm(\tau)$ and ``twist''
$\tau^\pm(\tau)$
fields that change
the boundary conditions as one would go from $\sigma=0$ to
$\sigma=\pi$. As done in \cite{Itoyama}, we will fix $\tau_1,\tau_2,\tau_3$
at $0,-\infty,-1$ respectively. We now evaluate several tree-level amplitudes
in the open and mixed sectors (in the next section).

(I) $A_{ooo^\prime}$

We first evaluate the 3-point function $A_{ooo^\prime}$
involving two $p-p^\prime$ open strings denoted by $o$ each, 
and one $p^\prime-p^\prime$ open string denoted by $o^\prime$.
Now, $A_{ooo^\prime}$ can be read off from equation
(4.25) of \cite{Itoyama} and is given by 
(having used the Jacobian for gauge-fixing the super-M\"{o}bius
symmetry (See \cite{Mar})):
\begin{eqnarray}
\label{eq:3point}
& & A_{ooo^\prime}
={(\tau_1-\tau_2)(\tau_2-\tau_3)(\tau_3-\tau_1)\over(\tau_1-\tau_2)^2}
\int d\theta_3 \nonumber\\
& & \langle0|
:\sigma^+(\tau_1)\tau^+(\tau_1)
e^{i(k\cdot{\bar x}+{\bar k}\cdot x)^{1{\bar1}}(\tau_1)}:
:\sigma^-(\tau_2)\tau^-(\tau_2)
e^{i(k\cdot{\bar x}+{\bar k}\cdot x)^{1{\bar 1}}(\tau_2)}:\nonumber\\
& & \times\int d\eta_1:e^{i(k\cdot{\bar x}+{\bar k}\cdot x)(\tau_3)
+ik\eta_1(D_L+D_R)(k\cdot{\bar x}-{\bar k}\cdot
  x)(\tau_3)}:|0\rangle\nonumber\\
& & \sim i\delta(\sum_{a=1}^3k_{a1})\delta(\sum_{b=1}^3k_{b{\bar 1}})
\sqrt{{\alpha^\prime\over2}}(k_2-k_1)\pdot k_3 e^{C_3(\nu)} 
e^{{i\over2}\Theta^{ij}k_{1i}k_{2j}}
\end{eqnarray}
(See (\ref{eq:reldefs}) for definition of $C_3(\nu)$).
The vacuum is a tensor product of the 
vacuum corresponding to the usual SL(2,{\bf R})-
invariant vacuum for directions $1, {\bar 1}$, and the vacuum
corresponding to the uncommon directions $2, {\bar 2}$ that is the
analog of the ``oscillator vacuum'' of \cite{Itoyama}.
From (\ref{eq:3point}), one sees that $A_{ooo^\prime}$ vanishes
on-shell. This is analogous to the similar result in \cite{Ademollo}.
Note that (\ref{eq:3point}) for off-shell scalars,
is written entirely in terms of
the $1-{\bar 1}$ subspace of the target space.

(II) $A_{ooo^\prime o^\prime}$

Next, we evalate the 4-point function $A_{ooo^\prime o^\prime}$
involving two $p-p^\prime$ open string and two $p^\prime-p^\prime$ 
open string vertex operators. The four-point function, defined as:
\begin{eqnarray}
\label{eq:4pointdef}
& & A_{ooo^\prime o^\prime}
={(\tau_1-\tau_2)(\tau_2-\tau_3)(\tau_3-\tau_1)\over(\tau_1-\tau_2)^2}
\int d\tau_4d\theta_3 d\theta_4 \nonumber\\
& & \langle0|
:\sigma^+(\tau_1)\tau^+(\tau_1)
e^{i(k\cdot{\bar x}+{\bar k}\cdot x)^{1{\bar1}}(\tau_1)}:
:\sigma^-(\tau_2)\tau^-(\tau_2)
e^{i(k\cdot{\bar x}+{\bar k}\cdot x)^{1{\bar 1}}(\tau_2)}:\nonumber\\
& & \times\int d\eta_1 :e^{i(k\cdot{\bar x}+{\bar k}\cdot x)(\tau_3)
+ik\eta_1(D_L+D_R)(k\cdot{\bar x}-{\bar k}\cdot
  x)(\tau_3)}:\nonumber\\
& &
\times\int  d\eta_2:e^{i(k\cdot{\bar x}+{\bar k}\cdot x)(\tau_4)
+ik\eta_2(D_L+D_R)(k\cdot{\bar x}-{\bar k}\cdot
  x)(\tau_4)}:|0\rangle,
\end{eqnarray}
can be read off from equation (4.27) of  \cite{Itoyama} after identification
of the polarization vector in the vector vertex operator $\zeta_\mu$
in \cite{Itoyama} with $ik_\mu$, and is given in (\ref{eq:A4}) 
in Appendix A.  One can extract the pole structure 
of (\ref{eq:A4}), as in \cite{Itoyama}, by evaluating in 
the Seiberg-Witten(SW) low energy limit, the integral 
around $x=0$ ($\int_0^\delta$) corresponding to the
$t-$channel process and around $x=1$ ($\int_{1-\delta}^1$)
corresponding to the $s$-channel exchange. As (almost) 
massless particle-exchange will dominate the contributions of various 
states to the above four-point function, we have 
to find the almost massless poles from the above expression.

The following observations are useful.
(a) Using $L_0=\alpha^\prime G^{1{\bar 1}}|k_1|^2 + {\nu\over2}=0$ for
$N=2$ theories, one gets $\alpha^\prime m_o^2={\nu\over2}(>0)\sim O(1)$,
(b) $\alpha^\prime\rightarrow \sqrt{\epsilon},
\ b_{2{\bar 2}}\rightarrow1/\sqrt{\epsilon}$, and $\nu\equiv1 
+ O(\sqrt{\epsilon})$.

The integral $\int_0^\delta$ gives terms of the type
${\alpha^\prime O(1)\over{\alpha^\prime t + O(1)}}$ and
${O(1)\over{\alpha^\prime t + O(1)}}$. One sees there are no terms of 
the type ${1\over{\alpha^\prime t + O(\sqrt{\epsilon})}}$, 
which would have corresponded to an almost massles pole. Hence, in the 
SW low energy limit, only
$\lim_{\alpha^\prime\rightarrow0}{O(1)\over{\alpha^\prime t +
    O(1)}}\equiv O(1)$ terms survive. These $O(1)$ terms are local.

As the $s$-channel corresponds to exchange of $p^\prime-p^\prime$ open string
that has only a massless scalar in its spectrum, the result can 
be read off from equation (5.11) of \cite{Itoyama}, and is local:
\begin{eqnarray}
A_4&&\sim
\frac{1}{2s}\left[  \left\{
   (k_{2}-k_{1})\pdot k_{3}-ik_{4}\ppdot  Jk_{3}
  \right\} k_{3}\prdot k_{4}\right.\nonumber\\
&&\hspace{4em}-k_{4}\prdot k_{3} \left\{
       (k_{2}-k_{1})\pdot k_{4}
     {}-ik_{3}\ppdot Jk_{4} \right\}\nonumber\\
&&\hspace{4em} -2 \nu
    \left(k_{3}\peke k_{4}\right)_{2}k_{3}\prdot k_{4}
   {}-2k_{3}\prdot k_{4} \nu
     \left(k_{3}\peke k_{4}\right)_{2}\nonumber\\
&&\hspace{4em}\left.
   + \left\{-t+m_{o}^{2} +k_{3}\ppdot k_{4}
     {}-(1-2\nu)
       \left(k_{3}\peke k_{4}\right)_{2}
    \right\}    k_{3}\prdot k_{4}
\right]\nonumber\\
&&\hspace{1.5em}\times \exp\left[2\alpha^{\prime}
     \left(k_{3}\kuroten k_{4}\right)_{2}
     \left\{\gamma+\frac{1}{2}\left(\mbox{\boldmath$\psi$}(\nu)
            +\mbox{\boldmath$\psi$}(1-\nu)\right)\right\}
         \right]\nonumber\\
&&+\left(k_{3}\leftrightarrow k_{4}\right),
\label{eq:nearx=1}
\end{eqnarray}
which we see is local. In (\ref{eq:nearx=1}) 
\begin{eqnarray}
 & & s\equiv -(k_{3}+k_{4})\prdot(k_{3}+k_{4})
  =-2 k_{3}\prdot k_{4}~;\nonumber\\
& &   J \equiv\left( {J_{\mu}}^{\rho}\right)\equiv
  \left( \begin{array}{cc}
0 & 0 \\
0 & i\sigma_2 \\
\end{array}\right).
\end{eqnarray}

To see if one is able to generate the expression for $A_{ooo^\prime o^\prime}$
from two 3-point functions, given that $A_{ooo}$ vanishes 
on-shell, one will have to evaluate mixed 3-point functions
corresponding to scattering of a graviton from a $p-p^\prime$ open
string - $A_{ooc}$,
as well as scattering of a graviton from a $p^\prime-p^\prime$ open
string - $A_{o^\prime o^\prime c}$. We discuss this in the next section.

\section{The mixed sector}

We now consider the mixed 3-point functions  involving two $p-p^\prime$ 
or $p^\prime-p^\prime$ open strings and a closed string, 
$A_{ooc}$ and $A_{o^\prime o^\prime c}$, respectively. 
Even though an exact answer can be obtained, we 
work in the Seiberg-Witten low energy limit followed by 
infinite noncommutativity limit eventually, as only then
can we get an answer that has no contact-term divergences.

(I) $A_{ooc}$

One can show that the following vertex operator written in term of
$N=1$ notations, 
reproduces the $N=2$ vertex operator for closed strings.
\begin{equation}
\label{eq:clN=1vop}
V^{\rm int}_c
=\int\int d\eta_1 d\eta_2\int\int d\theta_L d\theta_R
 Exp\biggl[E_I X^I + {\bar E}_{\bar I}{\bar X}^{\bar I}\biggr],
\end{equation}
where
\begin{eqnarray}
\label{eq:defsEs}
& & E_I\equiv i\sqrt{{\alpha^\prime\over2}}
l_I + il_I(\eta_1 D_L+\eta_2D_R);\nonumber\\
& & {\bar E}_{\bar I}\equiv i\sqrt{{\alpha^\prime\over2}}
{\bar l}_{\bar I} -i{\bar l}_{\bar I}(\eta_1D_L+\eta_2D_R).
\end{eqnarray}

For calculating self-contractions for the third closed vertex operator,
that needs to be done to go from the $SL(2,{\bf R})$-normal ordering
$:\ :$, to oscillator-normal ordering $::\ ::$, one has to evaluate:
$Exp\biggl(E_I{\bar E}_{\bar I}G^{{\rm sub}\ I{\bar I}}\biggr)$, where
$G^{{\rm sub}}$ is the subtracted 2-point function of \cite{Itoyama}
(also defined in appendix B).  This calculation 
of self-contraction is done in appendix B.
As the closed string metric $g^{I{\bar I}}$, the open-string metric
$G^{I{\bar I}}$ and the noncommutativity parameter 
${\Theta^{I{\bar I}}\over{\pi\alpha^\prime}}$ are 
given by ${\delta^{I{\bar I}}\over\epsilon}$,
${2\delta^{I{\bar I}}\over{\epsilon(1+b_I^2)}}$ and
${\delta^{I{\bar I}}b_I\over{\epsilon(1+b_I^2)}}$, one sees the
explicit appearance of  all three, in particular the closed-string
metric,  in (\ref{eq:selfcont1}). 

In the superspace formalism, for calculating $A_{ooc}$,
one has to evaluate:
\begin{eqnarray}
\label{eq:Aoocdef}
& &  
\int d(Re z_3)\int\int d\theta_Ld\theta_R\int\int d\theta_1d\theta_2\int\int d\eta_1d\eta_2
\nonumber\\
& & \langle 0|:\sigma^+(\tau_1)\tau^+(\tau_1)
\theta_1Exp[ik_1X^1+i{\bar k}_{\bar 1}{\bar X}^{\bar 1}](\tau_1,\theta_1):
:\sigma^-(\tau_2)\tau^-(\tau_2)
\theta_2Exp[iq_1X^1+i{\bar q}_{\bar 1}
{\bar X}^{\bar 1}](\tau_2,\theta_2):\nonumber\\
& & \times:Exp[E_1X^1+{\bar E}_{\bar 1}X^{\bar 1}
+E_2X^2+{\bar E}_{\bar 2}X^{\bar 2}]
(z_3,{\bar z}_3;\theta_L\theta_R):
|0\rangle.
\end{eqnarray}
Then, using $G^{\rm sub}$ to go from the SL(2,{\bf R})-invariant vacuum
$|0\rangle$ to the oscillator vacumm $|\sigma,s\rangle$,
in the $2,{\bar 2}$-subspace for the closed-string vertex operator, the 
above gives:
\begin{eqnarray}
\label{eq:Aoocdef21}
& & 
\int d(Re z_3)
\int\int d\theta_Ld\theta_R
\int\int d\theta_1d\theta_2\int\int d\eta_1d\eta_2\nonumber\\
& & \langle 0|:\theta_1
Exp[ik_1X^1+i{\bar k}_{\bar 1}{\bar X}^{\bar 1}](\tau_1,\theta_1):
:\theta_2Exp[iq_1X^1+i{\bar q}_{\bar 1}
{\bar X}^{\bar 1}](\tau_2,\theta_2):\nonumber\\
& & \times :Exp[E_1X^1+{\bar E}_{\bar 1}{\bar X}^{\bar 1}]
(z_3,{\bar z_3};\theta_L,\theta_R):|0\rangle\nonumber\\
& & \times Exp[E_I{\bar E}_{\bar I}G^{I{\bar I}}_{\rm sub}]
(z_3,{\bar z_3};\theta_L,\theta_R)
\times\langle
\sigma, s|:: Exp[E_2X^2+{\bar E}_{\bar 2}
{\bar X}^{\bar 2}](z_3,{\bar z_3};\theta_L,\theta_R)
::|\sigma, s\rangle,\nonumber\\
\end{eqnarray}
which is the same as:
\begin{eqnarray}
\label{eq:Aoocdef2}
& &
\int d(Re z_3)
\int\int d
\theta_Ld\theta_R\int\int d\theta_1d\theta_2\int\int d\eta_1d\eta_2
\nonumber\\
& & \langle 0|:\theta_1Exp[ik_1X^1+i{\bar k}_{\bar 1}{\bar X}^{\bar 1}]
(\tau_1,\theta_1):
:\theta_2Exp[iq_1X^1+i{\bar q}_{\bar 1}{\bar X}^{\bar
  1}](\tau_2,\theta_2):
\nonumber\\
& & \times :Exp[E_1X^1+{\bar E}_{\bar 1}{\bar X}^{\bar 1}]
(z_3,{\bar z_3};\theta_L,\theta_R):|0\rangle\nonumber\\
& & \times Exp[E_I{\bar E}_{\bar I}G^{I{\bar I}}_{\rm sub}]
(z_3,{\bar z_3};\theta_L,\theta_R).
\end{eqnarray}
Equation (\ref{eq:Aoocdef2})  of the form
\begin{eqnarray}
\label{eq:Aoocdef3}
& & 
\int d(Re z_3)
\int\int d\theta_Ld\theta_R\int\int d\theta_1d\theta_2\int\int d\eta_1d\eta_2
\rho e^{\eta_1\theta_L\alpha_1+\eta_2\theta_R\beta_1}
Exp[E_I{\bar E}_{\bar I}G^{I{\bar I}{\rm sub}}]
\nonumber\\
& & =\int d(Re z_3)\rho e^{\gamma_1}
\biggl(\delta_2+\gamma_2\delta_1
-(\alpha_1 [{\Theta^{1{\bar 1}}
\over{\pi\alpha^\prime}}] +\alpha_2 {\delta(0)\over\epsilon}
+\alpha_3[{\Theta^{2{\bar2}}
\over\pi\alpha^\prime}]+\alpha_4)
(\beta_1 [{\Theta^{1{\bar 1}}
\over{\pi\alpha^\prime}}]+
\beta_2{\delta(0)\over\epsilon}+\beta_3[{\Theta^{2{\bar2}}
\over\pi\alpha^\prime}]+\beta_4)\nonumber\\
& & 
-\alpha_5(\omega_1[{\Theta^{2{\bar2}}
\over\pi\alpha^\prime}]+\omega_2)\biggr),\nonumber\\
\end{eqnarray} 
where $\rho,\gamma_is,\alpha_is,\beta_is,\omega_is$ are defined
in (\ref{eq:DEFs}).
We will now work in the infinite noncommutativity limit in which one can
drop the $\delta(0)$-dependent terms
that appear as additive terms in (\ref{eq:Aoocdef3}),
relative to the
$\alpha_i[{\Theta^{1(2){\bar 1}({\bar 2})}\over{\pi\alpha^\prime}}]
\beta_j [{\Theta^{1(2){\bar 1}({\bar 2})}\over{\pi\alpha^\prime}}]
\propto\Theta^2$ terms in
(\ref{eq:Aoocdef3}). This can be seen more explicitly by choosing
the representation $\delta(x)
=\lim_{\epsilon^\prime\rightarrow0}
{\epsilon^\prime\over(x^2+{\epsilon^\prime}^2)}$. Thus, for
$\epsilon^\prime\sim\epsilon$,
${\delta(0)\over\epsilon}\sim{1\over\epsilon^2}$. Hence,
in the infinite noncommutativity limit, if one assumes that
$\Theta^{1{\bar 1}}$ and/or $\Theta^{2{\bar 2}}
\rightarrow\infty$ as ${1\over\epsilon^2}$,
this justifies dropping the $(\delta(0))^n, n=1,2$ terms relative to the
$\alpha_i[{\Theta^{1(2){\bar 1}({\bar 2})}\over{\pi\alpha^\prime}}]
\beta_j [{\Theta^{1(2){\bar 1}({\bar 2})}\over{\pi\alpha^\prime}}]
\propto{\Theta^2\over\epsilon}\sim{1\over\epsilon^5}$ terms. The
terms proportional to $\delta^\prime(0)$ can be dropped as
$\delta^\prime(0)\equiv\int_{-\infty}^\infty \delta^\prime(x)\delta(x)=0$
as $\delta(x)$ is an even function and $\delta^\prime(x)$
is an odd function.
Also, as the step function $\Theta(x)$ is related to the sign function
$\epsilon(x)$ by the relation
${1\over2}\epsilon(x)=\Theta(x)-{1\over2}$,
hence using the integral representation for $\epsilon(x)=
{1\over{2\pi i}}\int_{-\infty}^\infty 
d\alpha({e^{ix\alpha}\over{\alpha-i\epsilon}}
-{e^{-ix\alpha}\over{\alpha-i\epsilon}})$, we see that
$\lim_{x\rightarrow0}
\epsilon(x)=0$, implying $\lim_{x\rightarrow0}\Theta(x)={1\over2}$.
Now, in (\ref{eq:Aoocdef3}) and (\ref{eq:DEFs}), we have assumed that one has
fixed $Im z_3$, and hence one requires to integrate only over $Re z_3$.

The sum in equation (\ref{eq:DEFs}) for the expression for $\gamma_1$ should be
evaluated as follows. Consider the $cos$ -dependent and
independent  terms separately. The $cos$-independent terms
can be written as $-2 \gamma - (\Psi(1-\nu) +\Psi(\nu))$,
where $\gamma\equiv$ Euler number, and $\Psi\equiv\Gamma^\prime/\Gamma$.
Then use identity (A.5) of Itoyama's appendix, and one sees
that at $\nu=1$, the above is proportional to $\alpha^\prime b_{2{\bar
2}}\rightarrow\beta$.
$cos$-dependent terms can be evaluated by writing
$1-\nu=\sqrt{\epsilon}$.
One will get $1/\sqrt{\epsilon} - ln sin^2\epsilon_1$.
We demand that $\epsilon,\epsilon_1\rightarrow0$ in such in a way that:
\begin{equation}
\label{eq:eps0}
ln sin^2\epsilon_1 + {1\over{\sqrt{\epsilon}}}=0,
\end{equation}
which is reasonable as $ln sin^2\epsilon_1$ approaches $-\infty$, and
$1/\sqrt{\epsilon}$ approaches $+\infty$. Hence, the
$\alpha^\prime ln[sin^2(\epsilon_1)]$-type singularity is repaced by
$\alpha^\prime/\sqrt{\epsilon}\sim \beta$.
Now comes the point of evaluating the sum in $\gamma_1$.
Take the Seiberg-Witten
low energy limit and hence take $1-\nu=\sqrt{\epsilon}$ .
The the sum becomes:
\begin{equation}
\label{eq:sumSWlt}
2\sum_{m=1}^\infty{cos[m\sigma_3]\over m}=-ln(sin^2{\sigma_3\over2}).
\end{equation}
For the evaluation of the integral, it is more 
convenient to evaluate the integral using $Re z_3$. One has to evaluate:
\begin{equation}
\label{eq:intRez_3}
\int_{-\infty}^\infty d Re z_3
(Re z_3 + i Im z_3)^{\lambda_4}(Re z_3 - i Im z_3)^{\lambda_5} e^{\lambda_1 ln(
1+cos\sigma_3)}
(1,\lambda_3 e^{-2\tau_3}).
\end{equation}
The above integral is evaluated in Appendix B.
We now consider three cases for infinite noncommutativity.

(a) \underline{$\Theta^{1,{\bar 1}}\rightarrow\infty$,
$\Theta^{2{\bar 2}}\equiv$ finite}

From (\ref{eq:DEFs})
and (\ref{eq:intresult}), one sees that in the Seiberg-Witten
low energy limit, $\lambda_1\sim O(1/\sqrt{\epsilon})$,
$\lambda_4-\lambda_5\sim2\lambda_4\sim-2\lambda_5
\equiv O(\Theta^{1{\bar1}}/\sqrt{\epsilon})
\equiv O({1\over\epsilon^{{5\over2}}})>>\lambda_1$.
Then, using
\begin{eqnarray}
\label{eq:simpshypergeoom1}
& & (a) Appell F_1(a,b_1,b_2;a;x,y)=Appell F_1 (1,b_1,b_2;1;x,y);\nonumber\\
& & (b) \ _2F_1(a,b;a;x)=\ _2F_1(1,b;1;x);\nonumber\\
& & (c) Appell F_1(1,a,b;1;-i,i)=(1-i)^{-b}(1+i)^{-a};\nonumber\\
& & (d) \ _2F_1(1,a;1;-1)=2^{-a},
\end{eqnarray}
one sees that one gets:
\begin{eqnarray}
\label{eq:finR}
& & \lambda_3\Gamma(2\lambda_1){e^{{-i\pi(\lambda_4-
\lambda_5)\over2}+{i\pi\lambda_5}}
\over{-4\lambda_4}} +(k\leftrightarrow q).\nonumber\\
& & \end{eqnarray}
Hence, one gets
a factor of ${e^{{{-i(\lambda_4-
\lambda_5)\pi}\over2}+i\pi\lambda_5}\over2\lambda_4}
\stackrel{\Theta^{1{\bar 1}}\rightarrow\infty}{\rightarrow}$
${e^{-i\pi(\lambda_4-\lambda_5)}\over{(\lambda_4-\lambda_5)}}$
$={e^{i\Theta^{1{\bar 1}}(k_1
{\bar l}_{\bar 1}-{\bar k}_{\bar 1}l_1)}\over{\Theta^{1{\bar 1}}
({\bar l}_{\bar 1}k_1 - {\bar k}_{\bar 1}l_1)}}$, which gives a generalized
star product \cite{LM} after Bose symmetrization
in the noncommutative $p^\prime-p^\prime$ $A_{ooc}$ amplitude.

In the $\Theta^{1{\bar 1}}\rightarrow\infty$ limit, and 
setting $(Im z_3)_0=1$, (\ref{eq:finR}) simplifies to give:
\begin{eqnarray}
\label{eq:1infinity}
& & A_{ooc}(\epsilon\rightarrow0,\Theta^{1{\bar 1}}\rightarrow\infty)
\sim \nonumber\\
& & {2\over\alpha^\prime} e^{{-2\pi|l_2|^2\beta\over\epsilon}}
\biggl(
\Theta^{1{\bar 1}}(k_1{\bar q}_{\bar 1}-{\bar k}_{\bar 1}q_1)
\biggr)^2
{sin[\Theta^{1{\bar 1}}(k_1{\bar q}_{\bar 1}-{\bar k}_{\bar 1}q_1)]\over
{\Theta^{1{\bar 1}}(k_1{\bar q}_{\bar 1}-{\bar k}_{\bar 1}q_1)}}
\Gamma({2|l_2|^2\alpha^\prime\over\epsilon}).
\end{eqnarray}
The above result consists of product of four factors - a gaussian damping factor
\cite{Itoyama,gdf}
``gdf"[$g^{-1}$]  analogous to the open sector but here it 
depends on the closed string metric, the factor  
``$c_L$" where (in the extreme noncommutativity limit),
$``c_L=-c_R"=\Theta^{1{\bar 1}}(k_1{\bar q}_{\bar 1} - {\bar k}_{\bar 1}q_1)$
(See \cite{JHEP1}), a generalized star product, and a kinematic factor
$\Gamma({2|l_2|^2\alpha^\prime\over\epsilon})$. 
Hence, $A_{ooc}$ also involves the 
closed-string metric in addition to the open-string metric.
The expression (\ref{eq:1infinity}) corresponds to an infinite series of local
interactions.

(b) \underline{$\Theta^{1{\bar 1}},\Theta^{2{\bar 2}}\rightarrow\infty$}

In this limit, one has to evaluate:
\begin{eqnarray}
\label{eq:12infinity1}
& & \int_{-\infty}^\infty d Re z_3 e^{\lambda_1 ln(1+cos\sigma_3)}
(Re z_3 + i Im z_3)^{\lambda_4}(Re z_3 - i Im z_3)^{\lambda_5} 
\Biggl( \alpha_1\beta_1 +\alpha_1\beta_3 +\beta_1\alpha_3 
+\alpha_3\beta_3\Biggr).\nonumber\\
& & 
\end{eqnarray}
Thus, one gets:
\begin{eqnarray}
\label{eq:12infinity}
& & A_{ooc}(\epsilon\rightarrow0,
\Theta^{1{\bar1},2{\bar2}}\rightarrow\infty)\sim\nonumber\\
& & 
e^{{-2\pi|l_2|^2\beta\over\epsilon}}
{\Gamma({2|l_2|^2\alpha^\prime\over\epsilon})\over\alpha^\prime}
{sin[\Theta^{1{\bar 1}}(k_1{\bar q}_{\bar 1}-{\bar k}_{\bar 1}q_1)]\over
{\Theta^{1{\bar 1}}(k_1{\bar q}_{\bar 1}-{\bar k}_{\bar 1}q_1)}}
\biggl[-2\alpha^\prime\biggl(
\Theta^{1{\bar 1}}(k_1{\bar q}_{\bar 1}-{\bar k}_{\bar 1}q_1)
\biggr)^2
+\biggl(\sqrt{{\alpha^\prime\over2}}
{4i|l_2|^2\over{Im z_3}}{\Theta^{2{\bar2}}
\over{\pi\alpha^\prime}}\biggr)^2\biggr].\nonumber\\
& & \end{eqnarray}
Again, one sees the appearance of a generalized product in the amplitude.
As in (\ref{eq:1infinity}), (\ref{eq:12infinity}) corresponds to an infinite
series of local terms.

(c) \underline{$\Theta^{2{\bar 2}}\rightarrow\infty, \Theta^{1{\bar 1}}\equiv$
finite}

One has to evaluate in this limit: 
\begin{eqnarray}
\label{eq:2infinity1}
& & \int_{-\infty}^\infty d Re z_3 \
\alpha_3\beta_3
e^{\lambda_1 ln(1+cos\sigma_3)}
(Re z_3 + i Im z_3)^{\lambda_4}(Re z_3 - i Im z_3)^{\lambda_5}.
\end{eqnarray}
Hence, one gets:
\begin{eqnarray}
\label{eq:2infinity}
& & A(\epsilon\rightarrow0,\Theta^{2{\bar 2}}\rightarrow\infty)
\sim\nonumber\\
& & 
\sim 
{\biggl(\sqrt{{\alpha^\prime\over2}}
{4i|l_2|^2\over{Im z_3}}{\Theta^{2{\bar2}}
\over{\pi\alpha^\prime}}\biggr)^2\over
({|l_2|^2\alpha^\prime\over\epsilon})}
e^{{-2\pi|l_2|^2\beta\over\epsilon}}
2^{
\alpha^\prime G^{1{\bar 1}}
(k_1{\bar l}_{\bar 1}+{\bar k}_{\bar 1}l_1)
}\Biggl(\Gamma(-1-\alpha^\prime G^{1{\bar 1}}
(k_1{\bar l}_{\bar 1}+{\bar k}_{\bar 1}l_1))
cos[{3\Theta^{1{\bar 1}}(k_1{\bar q}_{\bar 1}-{\bar k}_{\bar 1}q_1)\over2}]
\nonumber\\
& & + 
{\Gamma({2|l_2|^2\alpha^\prime\over\epsilon})
\Gamma(-
{|l_2|^2\alpha^\prime\over\epsilon})
\over{\Gamma({|l_2|^2\alpha^\prime\over\epsilon})}}\ _2F_1\biggl(
-{|l_2|^2\alpha^\prime\over\epsilon},
2+\alpha^\prime G^{1{\bar 1}}
(k_1{\bar l}_{\bar 1}+{\bar k}_{\bar 1}l_1),{|l_2|^2\alpha^\prime\over\epsilon}
;-1\biggr)
cos[\Theta^{1{\bar 1}}(k_1{\bar q}_{\bar 1}-{\bar k}_{\bar 1}q_1)]
\Biggr).\nonumber\\
& & 
\end{eqnarray}
One gets another generalized star product different from the one that 
appears in cases (a) and (b). By writing $\epsilon=\epsilon_1+i\epsilon_2$,
and then using Stirling's asymptotic expression for the gamma function, 
and also using $\lim_{x\rightarrow\infty}\ _2F_1(x,a;-x;-1)
=\ _2F_1(1,a;1;1)$,
one
sees that (\ref{eq:2infinity}) also corresponds to an infinite series of
local terms.

To calculate $A_{ooo^\prime o^\prime}$, one needs to calculate
$A_{o^\prime o^\prime c}$ in addition to $A_{ooc}$. 

(II) $A_{o^\prime o^\prime c}$

As the ``tachyon'' vertex operators for the $p^\prime-p^\prime$ open
strings are $e^{i(k\cdot{\bar x}+{\bar k}\cdot x)}$, where all four
space-time coordinates are included, hence, the vacuum relevant to
this amplitude is only the SL(2,{\bf R})-invariant vacuum. Hence, this
amplitude is the same as the one calculated in \cite{JHEP1}. The
result is:
\begin{equation}
\label{eq:Ao'o'c}
A_{o^\prime o^\prime c}= c_L c_R {sin(k\Theta q)\over(k\Theta q)},
\end{equation}
where $c_{L,R}$ are as defined in \cite{JHEP1}. 

It  does not seem plausible to be able to obtain the local $t$- and
$s-$channel results of
$A_{ooo^\prime o^\prime}$ for finite noncommutativity 
or in the extreme noncommutativity
limit using $A_{ooc}$ of (\ref{eq:1infinity}) or (\ref{eq:12infinity}) or
(\ref{eq:2infinity})
and $A_{o^\prime o^\prime c}$
of (\ref{eq:Ao'o'c}). Some of the possible field theory graphs in the
$t$ and $s$ channels are drawn in Fig.1. As $A_{ooo^\prime}$ vanishes,
graphs (a) and (e) in Fig.1 vanish.
One requires to evaluate  a four-point function with an internal 
$p-p^\prime$ open-string exchange. The non-local pieces of all allowed
field theory graphs should cancel and the graphs should 
possibly give (though not necessary) the local part of $A_{ooo^\prime o^\prime}$
as obtained from string theory. One has to remember to 
include the contribution from $\int_\delta^{1-\delta}$ 
to get the complete form of the local expression.
Alternatively, it is possible that loop graphs in the field theory 
are required to be evaluated for reproducing a tree-level string
amplitude.  

\section{Summary and discussion}

To summarize, we have evaluated (by mapping the $N=2$ vertex operators
and hence the amplitudes to their $N=1$ counterparts) 
the 3-point and 4-point amplitudes involving two $p-p^\prime$ open strings 
and one or two $p^\prime-p^\prime$ open strings. While 
the former was found to vanish on-shell 
({\it suggesting the possibility that the noncommutative 
$N=2$ $p-p^\prime$ system in two complex dimensions
in SW low energy limit\footnote{The SW limit needs to be taken as even though
the 3-point function vanishes without having to take this limit, the 4-point 
function is local only after having taken the SW low energy limit. The
4-point function of (non)commutative $N=2$ $p^\prime-p^\prime$ in two
real dimensions, is local \cite{Ademollo}.} is related to a (non)commutative 
$N=2$ $p^\prime-p^\prime$ system in two real dimensions}), 
the latter in the Seiberg-Witten low-energy limit,
gave a {\it local result} for the $t$-channel and $s$-channel 
processes indicating {\it the possible topological 
nature (in the SW low energy limit) of the theory}.
We also evaluate the mixed 3-point function involving a closed
string and two $p-p^\prime$ or $p^\prime-p^\prime$ open strings.
While for finite noncommutativity, for the former, one 
obtains $\delta(0)$- and $\biggl(\delta(0)\biggr)^2$-type singularities 
(tree-level amplitudes in light cone 
$N=1$ string field theory are known to be singular which hence
require local divergent contact interactions as counter terms (to give finite
results) \cite{Greensite:1988hm} whose existence was argued earlier from
the super-Poincare algebra \cite{Greensite:1987sm}) 
while evaluating self-contractions for the closed-string vertex operator, these 
singularities can be avoided by taking the infinite noncommutativity limit in 
a suitable way. We consider three infinite noncommutativity 
limits (alongwith the Seiberg-Witten low energy limit): (a) 
$\Theta^{1{\bar 1}}\rightarrow\infty$ and $\Theta^{2{\bar 2}}\equiv$ finite, 
(b) $\Theta^{1{\bar 1}},\Theta^{2{\bar 2}}\rightarrow\infty$, and 
(c) $\Theta^{2{\bar 2}}\rightarrow\infty$ and $\Theta^{1{\bar 1}}\equiv$ finite.
Cases (a) (intriguing similar to the $A_{ooc}$ result of \cite{JHEP1})
and (b) give the commutative non-associative generalized star 
products involving ${sin(\partial\Theta\partial)\over(\partial\Theta\partial)}$,
and case (c) gives another commutative non-associative
generalized star product involving $cos(\partial\Theta\partial)$. All 
three cases involve a gaussian damping factor, similar to the $N=1$ calculations
in \cite{Itoyama,gdf}. The mixed 3-point function 
involving two $p^\prime-p^\prime$ open strings is identical to the corresponding
3-point function calculated for noncommutative $p^\prime-p^\prime$
system. It will be interesting to work on the field theory that
would reproduce the local 4-point function $A_{ooo^\prime o^\prime}$.

\section*{Acknowledgement}

We would like to thank the Abdus Salam ICTP, Trieste, Italy
for its hospitality where part of this work was done. We also
thank
Kamal L. Panigrahi, and especially Alok Kumar, for useful discussions.

\appendix
\section*{Appendix A}
\setcounter{equation}{0}
\seceqaa

We give below the result for the 4-point function $A_{ooo^\prime o^\prime}$
obtained after the identification of $\zeta=ik$ in the corresponding result
in \cite{Itoyama}.
\begin{eqnarray}
 \lefteqn{  A_{ooo^\prime o^\prime}\sim-
\delta(\sum_{a=1}^4k_{a1})\delta(\sum_{b=1}^4k_{b{\bar 1}})  
  \int^1_0 dx x^{-\alpha^{\prime}t+\alpha^{\prime}m_{o}^{2}}
  (1-x)^{2\alpha^{\prime} k_{3}\pdot k_{4}}
  \exp \Big({\cal C}_{3}(\nu)+{\cal C}_{4}(\nu)
              + ({\rm NC}) \Big) }
\nonumber\\
&&\times \exp \left[ -\alpha^{\prime}
     \left\{ \left(k_{3}\kuroten k_{4}+k_{3}\peke k_{4}\right)_{2}
          {\cal H}\left(\nu;\frac{1}{x}\right)
     +\left(k_{3}\kuroten k_{4}-k_{3}\peke k_{4}\right)_{2}
          {\cal H}\left(\nu;x \right)\right\}\right]
\nonumber\\
&&\times \Bigg[ \frac{1}{(1-x)^{2}}
   k_{3}\pdot k_{4}
   \left(1-2\alpha^{\prime}k_{3}\pdot k_{4}\right)
\nonumber\\
&&\hspace{0.5em}
   +\frac{\alpha^{\prime}}{2}  \frac{1}{x} \left\{\left[
      (k_{2}-k_{1})\pdot k_{3}  \right]
    {}- k_{4}\pdot k_{3}   \right\}
\left\{\left[(k_{2}-k_{1})\pdot k_{4}
    {}\right]+k_{3}\pdot k_{4}\right\}
\nonumber\\
&&\hspace{0.5em}+\alpha^{\prime} \frac{1}{1-x}\left\{
   \left[
       (k_{2}-k_{1})\pdot k_{3}\right]
   k_{3}\pdot k_{4}
 {} -k_{4}\pdot k_{3}
      \left[(k_{2}-k_{1})\pdot k_{4}\right]
\right\}\nonumber\\
&&\hspace{0.5em}+\frac{x^{-\nu} }{(1-x)^{2}}
   \left\{ -\alpha^{\prime}
  \left(k_{3}\kuroten k_{4} + k_{3}\peke k_{4}\right)_{2}
              k_{3}\pdot k_{4}
  \right.\nonumber\\
  &&\hspace{8em}\left.
     +\left(\frac{1-\nu}{2}-\alpha^{\prime}k_{3}\pdot k_{4}\right)
        \left(k_{3}\kuroten k_{4}
               + k_{3}\peke k_{4}\right)_{2}
      \right\} \nonumber\\
&&\hspace{0.5em}+\frac{x^{\nu}}{(1-x)^{2}}
   \left\{ -\alpha^{\prime}
    \left(k_{3}\kuroten k_{4} - k_{3}\peke k_{4}\right)_{2}
              k_{3}\pdot k_{4}
   \right.\nonumber\\
   &&\hspace{8em}\left.
     +\left(\frac{1-\nu}{2}-\alpha^{\prime}k_{3}\pdot k_{4}\right)
       \left(k_{3}\kuroten k_{4}
              {}-k_{3}\peke k_{4}\right)_{2}
   \right\}  \nonumber\\
&&\hspace{0.5em} +\frac{\nu}{2}
    \frac{x^{-\nu+1}}{(1-x)^{2}}
    \left(k_{3}\kuroten k_{4}
          + k_{3}\peke k_{4}\right)_{2}
    +\frac{\nu}{2}
     \frac{x^{\nu-1}}{(1-x)^{2}}
     \left(k_{3}\kuroten k_{4}
           {}-k_{3}\peke k_{4}\right)_{2} \nonumber\\
&&\hspace{0.5em}-\frac{\alpha^{\prime}}{2}
   \frac{x^{-2\nu}}{(1-x)^{2}}
   \left( k_{3}\kuroten k_{4} + k_{3}\peke k_{4}\right)_{2}
   \left( k_{3}\kuroten k_{4}
         +k_{3}\peke k_{4}\right)_{2}
\nonumber\\
&&\hspace{0.5em}-\frac{\alpha^{\prime}}{2}
   \frac{x^{2\nu}}{(1-x)^{2}}
   \left( k_{3}\kuroten k_{4} - k_{3}\peke k_{4}\right)_{2}
   \left( k_{3}\kuroten k_{4}
         {}-k_{3}\peke k_{4}\right)_{2}
\nonumber\\
&&\hspace{0.5em}-\frac{\alpha^{\prime}}{2}
   \frac{1}{(1-x)^{2}}
   \left( k_{3}\kuroten k_{4} + k_{3}\peke k_{4}\right)_{2}
   \left( k_{3}\kuroten k_{4}
         {}-k_{3}\peke k_{4}\right)_{2}
\nonumber\\
&&\hspace{0.5em}-\frac{\alpha^{\prime}}{2}
   \frac{1}{(1-x)^{2}}
   \left( k_{3}\kuroten k_{4} - k_{3}\peke k_{4}\right)_{2}
   \left( k_{3}\kuroten k_{4}
          +k_{3}\peke k_{4}\right)_{2}
\nonumber\\
&&\hspace{0.5em}+\frac{\alpha^{\prime}}{2}
   \frac{ x^{-2\nu} }{ 1-x }
   \left(k_{4}\kuroten k_{3}-k_{4}\peke k_{3}\right)_{2}
   \left(k_{3}\kuroten k_{4}+k_{3}\peke k_{4}\right)_{2}
\nonumber\\
&&\hspace{0.5em}-\frac{\alpha^{\prime}}{2}
   \frac{ x^{2\nu-1} }{ 1-x }
   \left(k_{4}\kuroten k_{3} + k_{4}\peke k_{3}\right)_{2}
   \left(k_{3}\kuroten k_{4} - k_{3}\peke k_{4}\right)_{2}
\nonumber\\
&&\hspace{0.5em}+\frac{\alpha^{\prime}}{2}
    \frac{ x^{\nu-1} }{ 1-x }\left\{\left(
    \left[(k_{2}-k_{1})\pdot k_{3}\right]
    {}-k_{4}\pdot k_{3}    \right)
    \left( k_{3}\kuroten k_{4}-k_{3}\peke k_{4}\right)_{2}
 \right.\nonumber\\
   && \hspace{5em}\left.-\left(
     \left[(k_{2}-k_{1})\pdot k_{4}\right]
      +k_{3}\pdot k_{4}\right)
          \left(k_{4}\kuroten k_{3}+k_{4}\peke k_{3}\right)_{2}
     \right\}\nonumber\\
&&\hspace{0.5em}+\frac{\alpha^{\prime}}{2}
   \frac{ x^{-\nu} }{1-x}\left\{\left(
   \left[(k_{2}-k_{1})\pdot k_{3}\right]
      +k_{4}\pdot k_{3}\right)
  \left(k_{3}\kuroten k_{4}+k_{3}\peke k_{4}\right)_{2}
 \right.\nonumber\\
   &&\hspace{5em}-\left.\left(
  \left[(k_{2}-k_{1})\pdot k_{4}\right]
      {}-k_{3}\pdot k_{4}\right)
  \left(k_{4}\kuroten k_{3}-k_{4}\peke k_{3}\right)_{2}
  \right\}
\Bigg]~.
\label{eq:A4}
\end{eqnarray}
where
\begin{eqnarray}
\label{eq:reldefs}
& & ({\rm NC})\equiv \sum_{1\leq a<a^{\prime}\leq N}\frac{i}{2}
              \epsilon(x_{a}-x_{a^{\prime}})
          \sum_{i,j=1}^{p}\theta^{ij}k_{ai}k_{a^{\prime}j};\nonumber\\
          & & 
k \ppdot k
  =    2G^{2\overline{2}}\kappa_{2}\overline{\kappa}_{\overline{2}}~;\nonumber\\
& & \  \left(k \kuroten q\right)_{2}
\equiv G^{2\overline{2}}(k_{2}\overline{q}_{\overline{2}}
                  +\overline{k}_{\overline{2}}q_{2})~,
\quad
\left(k \peke q \right)_{2}=G^{2\overline{2}}(k_{2}\overline{q}_{\overline{2}}
        {}-\overline{k}_{\overline{2}}q_{2})~,\nonumber\\
& & 
C_a(\nu)\equiv 2\alpha^\prime G^{2{\bar 2}}|k_2|^2
     \left\{\gamma+\frac{1}{2}\left(\mbox{\boldmath$\psi$}(\nu)
+\mbox{\boldmath$\psi$}(1-\nu)\right)\right\}\nonumber\\
        & &  {\cal H}(\nu;z)=\left\{
        \begin{array}{ll}
         \displaystyle{\cal F} \left(1-\nu_{I};\frac{1}{z}\right)
              {}-\frac{\pi}{2} b_{I}
                 =\sum_{n=0}^{\infty}\frac{z^{-n-1+\nu_{I}}}{n+1-\nu_{I}}
                      {}-\frac{\pi}{2} b_{I}&
                           \mbox{ for $|z| > 1$}\\[3ex]
                            \displaystyle{\cal F} \left( \nu_{I};z \right)+\frac{\pi}{2}b_{I}
                              =\sum_{n=0}^{\infty}\frac{z^{n+\nu_{I}}}{n+\nu_{I}}
                                 +\frac{\pi}{2}b_{I} &
                                      \mbox{ for $|z| <1$}
                                      \end{array}
                                      \right. ~, 
\end{eqnarray}
${\cal F}$  defined in (\ref{eq:Fdef}) again.

\section*{Appendix B}
\setcounter{equation}{0}
\seceqbb

In this appendix, we discuss the self-contraction calculation for
the closed-string vertex operator. For this purpose, one starts with:
\begin{eqnarray}
\label{eq:selfcont1}
& & 
Exp\biggl(E_I{\bar E}_{\bar I}G^{{\rm sub}\ I{\bar I}}\biggr)
=\lim_
{3\rightarrow3^\prime}
Exp\Biggl[\biggl(-{\alpha^\prime\over2}l_I{\bar l}_{\bar I}
-\sqrt{{\alpha^\prime\over2}}l_I{\bar l}_{\bar
  I}\biggl[\eta_1(D_L-D_L^\prime) +\eta_2(D_R-D_R^\prime)
\biggr]\nonumber\\
& &+l_I{\bar l}_{\bar
  I}\eta_1\eta_2(D_LD^\prime_R+D_RD^\prime_L)\biggr)
G^{\rm sub}\ ^{I{\bar I}}(z_3,{\bar z}_3;\theta_L,\theta_R|z^\prime,{\bar
  z}^\prime_3;\theta_L^\prime,\theta_R^\prime)\Biggr]\nonumber\\
& & \equiv
Exp\biggl(\gamma_1+\theta_L\theta_R\gamma_2+\eta_1\theta_L
(\alpha_2
{\delta(0)\over\epsilon}
+\alpha_3[{\Theta^{2{\bar2}}
\over\pi\alpha^\prime}]+\alpha_4)
+\eta_1\theta_R\alpha_5+\eta_2\theta_L
(\omega_1[{\Theta^{2{\bar2}}
\over\pi\alpha^\prime}]+\omega_2)\nonumber\\
& & +\eta_2\theta_R(
\beta_2{\delta(0)\over\epsilon}+\beta_3[{\Theta^{2{\bar2}}
\over\pi\alpha^\prime}]+\beta_4)
+\eta_1\eta_2\delta_1 + \eta_1\eta_2\theta_L\theta_R\delta_2\biggr),
\end{eqnarray}
where $\rho,\alpha_i,\beta_j,\omega_k,\gamma_l,\delta_m$s are defined in 
(\ref{eq:DEFs}), and the $G^{\rm sub}$ is the subtracted
Green's function of \cite{Itoyama}.

Now, $G^{\rm sub}\equiv{\bf{\cal G}}-{\bf G}$, and using
notations of \cite{Itoyama}, the following results
(of Itoyama et al) are used in arriving at (\ref{eq:DEFs}):
\begin{eqnarray}
&&\mbox{\boldmath$\cal G$}^{I\overline{J}}
   \left({\bf z}_{1},\overline{\bf z}_{1}
         | {\bf z}_{2},\overline{\bf z}_{2}\right)
 \equiv
   \left\langle \sigma,s \right|{\cal R}
     {\bf X}^{I}({\bf z}_{1},\overline{\bf z}_{1})
      \overline{\bf X}^{\overline{J}}({\bf z}_{2},\overline{\bf z}_{2})
  \left| \sigma,s\right\rangle\nonumber\\
&&=\Theta(|z_{1}|-|z_{2}|)\frac{2\delta^{I\overline{J}}}{\varepsilon}
  \left[
  {\cal F}\left(1-\nu_{I} \, ;
                 \, \frac{z_{2}+\theta_{1}\theta_{2}}{z_{1}}\right)
 +{\cal F}\left(1-\nu_{I}\, ; \,
     \frac{\overline{z}_{2}+\overline{\theta}_{1}\overline{\theta_{2}}}
          {\overline{z_{1}}}\right)
    \right.\nonumber \\
 && \hspace{9em}\left.
 {}-{\cal F}\left(1-\nu_{I}\,;\,
     \frac{\overline{z}_{2}+\theta_{1}\overline{\theta_{2}}}{z_{1}}
            \right)
 {}-{\cal F}\left(1-\nu_{I};
     \frac{z_{2}+\overline{\theta}_{1}\theta_{2}}
          {\overline{z_{1}}}\right)\right]\nonumber\\
&& +\Theta(|z_{2}|-|z_{1}|)\frac{2\delta^{I\overline{J}}}{\varepsilon}
   \left[
  {\cal F}\left(\nu_{I} \, ;
                 \, \frac{z_{1}}{z_{2}+\theta_{1}\theta_{2}}\right)
 +{\cal F}\left(\nu_{I}\, ; \,
     \frac{\overline{z_{1}}}
          {\overline{z}_{2}+\overline{\theta}_{1}\overline{\theta_{2}}}
    \right)   \right.\nonumber \\
 && \hspace{9em}\left.
 {}-{\cal F}\left(\nu_{I}\,;\,
     \frac{z_{1}}{\overline{z}_{2}+\theta_{1}\overline{\theta_{2}}}
            \right)
 {}-{\cal F}\left(\nu_{I};
     \frac{\overline{z_{1}}}
          {z_{2}+\overline{\theta}_{1}\theta_{2}}  \right)\right]~,
\label{eq:propCIMM-1}
\end{eqnarray}
where $\Theta(x)$ is the step function,
${\cal F}(\nu\,;\,z)$ is defined as
\begin{equation}
 \label{eq:Fdef}
 {\cal F}(\nu\,;\, z) =\frac{z^{\nu}}{\nu}\ _2F_1(1,\nu;1+\nu;z)
 =\sum_{n=0}^{\infty}\frac{1}{n+\nu}z^{n+\nu}~,
\end{equation}
and $\ _2F_1(a,b;c;z)$ is the hypergeometric function, 
and
\begin{eqnarray}
\label{eq:SL2invpt}
& & {{\bf G}^{I{\bar J}}\left({\bf z}_{1},\overline{\bf z}_{1}|
                   {\bf z}_{2},\overline{\bf z}_{2}\right)
  \equiv \langle 0 |
             {\cal R}{\bf X}^{I}({\bf z}_{1},\overline{\bf z}_{1})
             {\bf X}^{{\bar J}}({\bf z}_{2},\overline{\bf z}_{2})
           |0\rangle
   }\nonumber\\
  &&=-g^{I{\bar J}} \ln (z_{1}-z_{2}-\theta_{1}\theta_{2})
                     (\overline{z}_{1}-\overline{z}_{2}
                        {}-\overline{\theta}_{1}\overline{\theta}_{2})
     +(g^{I{\bar J}}-2G^{I{\bar J}}) \ln (z_{1}-\overline{z}_{2}
                        {}-\theta_{1}\overline{\theta}_{2})
                        (\overline{z}_{1}-z_{2}
                        {}-\overline{\theta}_{1}\theta_{2})
    \nonumber\\
   && \hspace{1em}-2\frac{\theta^{I{\bar J}}}{2\pi\alpha^{\prime}}
      \ln \biggl(\frac{z_{1}-\overline{z}_{2}-\theta_{1}\overline{\theta}_{2}}
               {\overline{z}_{1}-z_{2}-\overline{\theta}_{1}\theta_{2}}\biggr).
\end{eqnarray}

We give below the definitions of $\alpha_i,\beta_j,\omega_k,\gamma_l,
\delta_m$ that figure in the self-contractions in $A_{ooc}$.
\begin{eqnarray}
\label{eq:DEFs}
& &\gamma_1
\equiv -{l_2\delta^{2{\bar 2}}{\bar l}_{\bar 2}\alpha^\prime\over2\epsilon}
\biggl[2\sum_{m=0}^\infty
\biggl({[1-cos\{(m+1-\nu)\sigma_3\}]
\over(m+1-\nu)}+{[1-cos\{(m+\nu)\sigma_3\}]\over(m+\nu)}\biggr)\nonumber\\
& & -2ln sin^2\sigma_3 +2ln sin^2\epsilon_1
+{4\over(1+B^2_{2{\bar 2}})}
ln(4Im z_3^2)-{4B_{2{\bar 2}}\pi\over(1+B^2_{2{\bar 2}})}\biggr]
\nonumber\\
& & \gamma_2\equiv -{l_2\delta^{2{\bar 2}}{\bar l}_{\bar 2}\over\epsilon}\biggl(
{2i\over{Im z_3}}\biggl[
cos(2\nu\sigma_3)-cos(2(1-\nu)\sigma_3) - {(1-B^2_{2{\bar 2}})\over(1+B^2_{2{\bar 2}})}
\biggr]\biggr);\nonumber\\
& & \delta_1\equiv {4l_2\delta^{2{\bar 2}}{\bar l}_{\bar
    2}\over\epsilon}
{sin(2\nu\phi)-sin[2(1-\nu)\phi]
+sin(2\phi\nu)\over{Im z_3}};\nonumber\\
& & \rho\equiv 
(-Re z_3 - i Im z_3)^{{\alpha^\prime\over2}[k_1{\bar l}_{\bar 1}
(G^{1{\bar 1}}-{\Theta^{1{\bar 1}}\over{\pi\alpha^\prime}})+
{\bar k}_{\bar 1}l_{1}
(G^{1{\bar 1}}+{\Theta^{1{\bar 1}}\over{\pi\alpha^\prime}})]}
(-Re z_3 + i Im z_3)^{{\alpha^\prime\over2}[k_1{\bar l}_{\bar 1}
(G^{1{\bar 1}}+{\Theta^{1{\bar 1}}\over{\pi\alpha^\prime}})+
{\bar k}_{\bar 1}l_{1}
(G^{1{\bar 1}}-{\Theta^{1{\bar 1}}\over{\pi\alpha^\prime}})]} ;\nonumber\\
& & 
\alpha_1\equiv
-\sqrt{2\alpha^\prime}\biggl(
(-
G^{1{\bar 1}}+{\Theta^{1{\bar 1}}\over{\pi\alpha^\prime}})k_1{\bar l}_{\bar 1}-
(G^{1{\bar 1}}+{\Theta^{1{\bar 1}}\over{\pi\alpha^\prime}})
{\bar k}_{\bar 1}l_1\biggr)
{1\over{-Re z_3-i Im z_3}};\nonumber\\
& & \beta_1\equiv-\sqrt{2\alpha^\prime}
\biggl(
(-G
^{1{\bar 1}}-{\Theta^{1{\bar 1}}\over{\pi\alpha^\prime}})k_1{\bar l}_{\bar 1}+
(-G^{1{\bar 1}}+{\Theta^{1{\bar 1}}\over{\pi\alpha^\prime}})
{\bar k}_{\bar 1}l_1\biggr)
{1\over{-Re z_3+i Im z_3}}.\nonumber\\
& & \alpha_3\equiv\sqrt{{\alpha^\prime\over2}}
{4i|l_2|^2\over{Im z_3}}{\Theta^{2{\bar2}}
\over{\pi\alpha^\prime}};\nonumber\\
& & \alpha_4\equiv\sqrt{{\alpha^\prime\over2}}
{|l_2|^2g^{2{\bar 2}}\over{(Im
    z_3)_0}}(e^{2i(\nu-1)\phi} - e^{-2i\nu\phi});\nonumber\\
& & \alpha_5\equiv-\sqrt{{\alpha^\prime\over2}}{2|l_2|^2\over{Im z_3}}
G^{2{\bar 2}};\nonumber\\
&& \omega_1\equiv\sqrt{{\alpha^\prime\over2}}
{4i|l_2|^2\over{Im z_3}}{\Theta^{2{\bar2}}
\over{\pi\alpha^\prime}};
\nonumber\\
& & \omega_2\equiv\sqrt{{\alpha^\prime\over2}}
{i|l_2|^2\over{Im z_3}}g^{2{\bar 2}}sin(2\nu\phi);\nonumber\\
& &\beta_3\equiv\sqrt{{\alpha^\prime\over2}}
{4i|l_2|^2\over{Im
    z_3}}{\Theta^{2{\bar2}}\over{\pi\alpha^\prime}};\nonumber\\
& & \beta_4\equiv-\sqrt{{\alpha^\prime\over2}}{|l_2|^2g^{2{\bar 2}}\over{Im
    z_3}}(e^{2i(1-\nu)\phi} - cos(2\nu\phi))
\end{eqnarray}

\section*{Appendix C}
\setcounter{equation}{0}
\seceqcc

We discuss below the evaluation of the integral 
(\ref{eq:intRez_3}) in $A_{ooc}$.

\begin{eqnarray}
\label{eq:intRez3}
& & \int_{-\infty}^\infty d Re z_3 
(Re z_3 + i Im z_3)^{\lambda_4}(Re z_3 - i Im z_3)^{\lambda_5} e^{\lambda_1 ln(1+cos\sigma_3)}
(1,\lambda_3 e^{-2\tau_3})\nonumber\\
& & =\int_{-\infty}^\infty 
d(Re z_3)\biggl[1
+{Re z_3\over\sqrt{(Re z_3)^2+(Im z_3)^2}}
\biggr]^{\lambda_1}\biggl(
1,{\lambda_3\over{(Re z_3)^2+(Im z_3)^2_0}}\biggr)\nonumber\\
& & \times(Re _3
+i (Im z_3)_0)^{\lambda_4}(Re z_3+i (Im z_3)_0)^{\lambda_5}\nonumber\\
& & 
= (Im z_3)_0^{1+\lambda_4+\lambda_5}
\int_{-{\pi\over2}}^{{\pi\over2}} d\theta sec^2\theta
(1+sin\theta)^{\lambda_1}
\biggl(e^{-i[
\lambda_4-\lambda_5]
\theta}cos^{-2-\lambda_4-\lambda_5}\theta
, {\lambda_3\over{(Im z_3)_0^2}}e^{-i[\lambda_4-\lambda_5]\theta}
cos^{-[\lambda_4+\lambda_5]}\theta\biggr)
\nonumber\\
& &
=2^{\lambda_1-[\lambda_4+\lambda_5]}e^{i\pi(\lambda_4-\lambda_5)\over2}
(Im
z_3)_0^{1+\lambda_4+\lambda_5}\int_0^{\pi\over2}d\theta\biggl(
{e^{-i\pi(\lambda_4-\lambda_5)\over2}\over4}
e^{i[\lambda_4-\lambda_5]\theta}
cos^{2\lambda_1-2-(\lambda_4+\lambda_5)}{\theta\over2}
sin^{-2-(\lambda_4+\lambda_5)}{\theta\over2}\nonumber\\
& & +{e^{i\pi(2\lambda_2\lambda_4-\lambda_5)\over2}\over4}
e^{-i[\lambda_2+\lambda_4-\lambda_5]\theta}
cos^{-2-(\lambda_4+\lambda_5)}{\theta\over2}
sin^{2\lambda_1+\lambda_4-\lambda_5)}{\theta\over2},\nonumber\\
& & 
{\lambda_3\over(Im z_3)_0^2}e^{-i\pi(\lambda_4-\lambda_5)\over2}
e^{i(\lambda_4-\lambda_5)\theta}cos^{2\lambda_1-(\lambda_4+\lambda_5)}
{\theta\over2} sin^{-(\lambda_4+\lambda_5)}{\theta\over2}\nonumber\\
& & +{\lambda_3\over(Im z_3)_0^2}e^{i\pi(\lambda_4-\lambda_5)\over2}
e^{-i(\lambda_4-\lambda_5)\theta}cos^{-(\lambda_4+\lambda_5)}
{\theta\over2} sin^{2\lambda_1-(\lambda_4+\lambda_5)}
{\theta\over2}\biggr)
\end{eqnarray}
The above integrals can be evaluated using mathematica. One needs the 
following:
\begin{eqnarray}
\label{eq:math1}
& & \int_0^{{\pi\over2}} cos^a({x\over2}) sin^b({x\over2}) e^{icx}
dx\nonumber\\
& & 
={i2^{1 - a}\over{a + b - 2c}}\biggl[e^{i(b-a)\pi/4}2^{- b}e^{ic\pi/2}
{\rm AppellF1}[-{a\over2} - {b\over2} + c, -a, -b; 
1 - {a\over2} - {b\over2} + c; -i, i] \nonumber\\
& & - {\biggl(\Gamma[1 + b]\Gamma[1 - {a\over2} - {b\over2} + c] 
\ _2F_1[-{a\over2} - {b\over2} + c, -a; 
1 - {a\over2} + {b\over2} + c; -1]\biggr)
\over \Gamma[1 - {a\over2} + {b\over2} + c]}\biggr]
.
\end{eqnarray}
 One gets the following:
\begin{eqnarray}
\label{eq:intresult}
& &
2^{\lambda_1-[\lambda_4+\lambda_5]}e^{i\pi(\lambda_4-\lambda_5)\over2}
(Im
z_3)_0^{1+\lambda_4+\lambda_5}\Biggl[
{ie^{-i\pi(\lambda_4-\lambda_5)\over2}
2^{3-2\lambda_1+(\lambda_4+\lambda_5)}\over{4[2\lambda_1-4-4\lambda_4-2
\lambda_2]}}\biggl[e^{-i\lambda_1\over2}2^{2+\lambda_4+\lambda_5}
e^{i\pi[\lambda_4-\lambda_5]\over2}\nonumber\\
& & 
{\rm AppellF1}\biggl({2+\lambda_4+\lambda_5-2\lambda_1\over2}
+{2+\lambda_4+\lambda_5\over2}+\lambda_4-\lambda_5,
\lambda_4+\lambda_5-2\lambda_1+2,\nonumber\\
& & 
2+\lambda_4+\lambda_5;1+{2+\lambda_4+\lambda_5-2\lambda_1\over2}+{2+\lambda_4+\lambda_5\over2}+\lambda_4-\lambda_5;-i,i\biggr)
\nonumber\\
& &
-{\Gamma(-1-\lambda_4-\lambda_5)\Gamma(1+{2+\lambda_4+\lambda_5-2\lambda_1\over2}+{2+\lambda_4+\lambda_5\over2}+\lambda_4-\lambda_5)
\over\Gamma(1+{2+\lambda_4+\lambda_5
-2\lambda_1\over2}-{2+\lambda_4+\lambda_5\over2}+\lambda_4-\lambda_5
)}
\nonumber\\
& & \times\ _2F_1\biggl({2+\lambda_4+\lambda_5-2\lambda_1\over2}
+{2+\lambda_4+\lambda_5\over2}+\lambda_4-\lambda_5,\lambda_4+\lambda_5+2-2\lambda_1;\nonumber\\
& &
1+{2+\lambda_4+\lambda_5-2\lambda_1\over2}-{2+\lambda_4+\lambda_5\over2}
+\lambda_4-\lambda_5;-1\biggr)
\biggr]
\nonumber\\
& &+{ie^{i\pi(\lambda_4-\lambda_5)\over2}
2^{3+(\lambda_4+\lambda_5)}\over{4[2\lambda_1-4-4\lambda_4-2
\lambda_2]}}\biggl[e^{i\lambda_1\over2}2^{2+\lambda_4+\lambda_5-2\lambda_1}
e^{i\pi[\lambda_4-\lambda_5]\over2}\nonumber\\
& & {\rm AppellF1}\biggl({2+\lambda_4+\lambda_5-2\lambda_1\over2}
+{2+\lambda_4+\lambda_5\over2}-\lambda_4-\lambda_5,
\lambda_4+\lambda_5+2,\nonumber\\
& & 
2+\lambda_4+\lambda_5-2\lambda_1;1
+{2+\lambda_4+\lambda_5-2\lambda_1\over2}
+{2+\lambda_4+\lambda_5\over2}-(\lambda_4-\lambda_5);-i,i\biggr)
\nonumber\\
& &
-{\Gamma(2\lambda_1-1-\lambda_4-\lambda_5)\Gamma(1+{2+\lambda_4
+\lambda_5-2\lambda_1\over2}+{2+\lambda_4+\lambda_5\over2}
-(\lambda_4-\lambda_5))\over\Gamma(1+{2+\lambda_4+\lambda_5
-2\lambda_1\over2}-{2+\lambda_4+\lambda_5\over2}+\lambda_4-\lambda_5
)}
\nonumber\\
& & \times\ _2F_1\biggl({2+\lambda_4+\lambda_5-2\lambda_1\over2}
+{2+\lambda_4+\lambda_5\over2}-(\lambda_4-\lambda_5),
\lambda_4+\lambda_5+2;\nonumber\\
& &
1+{2+\lambda_4+\lambda_5-2\lambda_1\over2}-{2+\lambda_4+\lambda_5\over2}
-(\lambda_4-\lambda_5)
;-1\biggr)
\biggr],
\nonumber\\
& & 
{\lambda_3\over(Im z_3)_0^2}{ie^{-i\pi(\lambda_4-\lambda_5)\over2}
2^{1-2\lambda_1+(\lambda_4+\lambda_5)}\over{4[2\lambda_1-4\lambda_4]}}
\biggl(
e^{-i\lambda_1\pi\over2}2^{\lambda_4+\lambda_5}e^{i\pi(\lambda_4-\lambda_5)
\over2}\nonumber\\
& & 
\times{\rm AppellF_1}\biggl(-\lambda_1+2\lambda_4,-2\lambda_1
+(\lambda_4+\lambda_5);(\lambda_4+\lambda_5);-\lambda_1+1+2\lambda_4
;-i,i\biggr)\nonumber\\
& & -{\Gamma(1-\lambda_4-\lambda_5)\Gamma(1-\lambda_1+2\lambda_4)
\ _2F_1\biggl(-\lambda_1+2\lambda_4,-2\lambda_1+\lambda_4+\lambda_5;
1+\lambda_4-\lambda_5-\lambda_1;-1\biggr)
\over\Gamma(1+\lambda_4-\lambda_5-\lambda_1)}
\biggr)
\nonumber\\
& &+{\lambda_3\over(Im z_3)_0^2}{ie^{+i\pi(\lambda_4-\lambda_5)\over2}
2^{1+(\lambda_4+\lambda_5)}\over{4[2\lambda_1-4\lambda_4]}}
\biggl(
e^{i\lambda_1\pi\over2}2^{\lambda_4+\lambda_5-2\lambda_1}e^{-i\pi(\lambda_4
-\lambda_5)\over2}
\nonumber\\
& & 
\times{\rm AppellF_1}\biggl(-\lambda_1+2\lambda_5,(\lambda_4+\lambda_5),
-2\lambda_1+(\lambda_4+\lambda_5); 1 - 2\lambda_1+2\lambda_5
;-i,i\biggr)\nonumber\\
& & -{\Gamma(1-\lambda_4-\lambda_5)\Gamma(1-\lambda_1+2\lambda_4)
\ _2F_1\biggl(-\lambda_1+2\lambda_4,-2\lambda_1+\lambda_4+\lambda_5;
1+\lambda_4-\lambda_5-\lambda_1;-1\biggr)
\over\Gamma(1-
\lambda_4+\lambda_5-\lambda_1)}
\biggr) 
\Biggr]\nonumber\\
& & \end{eqnarray}

\clearpage
\begin{figure}[htbp]
\centerline{\mbox{\psfig{file=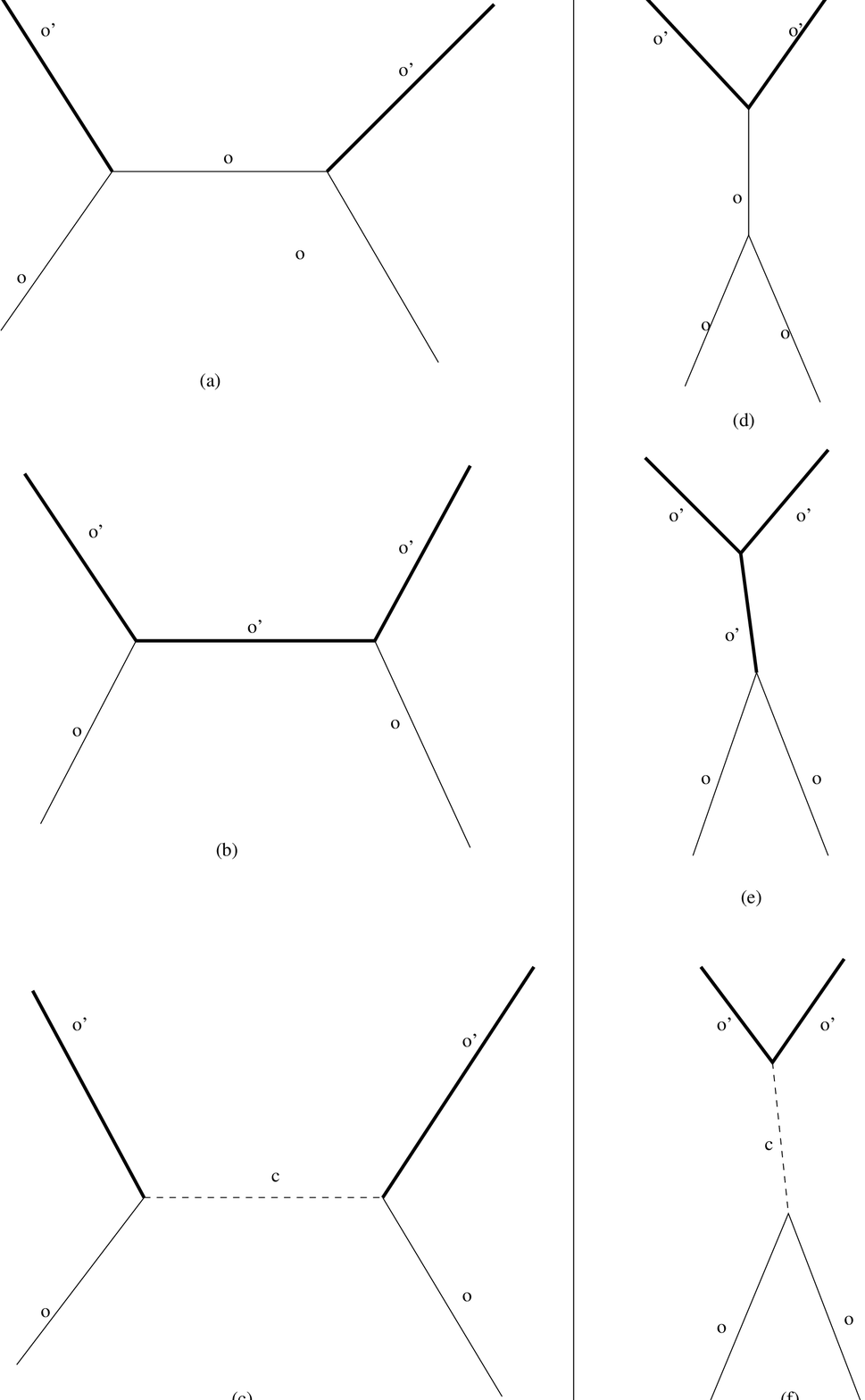,width=0.8\textwidth}}}
\caption{Some field theory graphs in the t and s channels; thin line is
  $p-p^\prime$
open-string scalar $o$, thick line is $p^\prime-p^\prime$ open-string
scalar $o'$, and dashed line is closed-string scalar $c$}
\end{figure}

\begin{thebibliography}{99}
\bibitem{Itoyama} 
B.Chen, H. Itoyama, T. Matsuo and K. Murakami, 
{\it Worldsheet and Spacetime Properties 
of p-p' System with B Field and Noncommutative Geometry}, 
Nucl. Phys.
{\bf B593}, 505, 2001 [hep-th/0005283].
\bibitem{SW} N.Seiberg, E.Witten,
{\it String Theory and Noncommutative Geometry},
JHEP 9909:032, (1999), 
[hep-th/9908142].
\bibitem{OV} H.~Ooguri and C.~Vafa,
{\it Selfduality And N=2 String Magic},
Mod. Phys. Lett. A {\bf 5} (1990) 1389, {\it Geometry of N=2 strings},
Nucl.\ Phys.\ B {\bf 361}, 469 (1991).
\bibitem{Mar} N. Marcus, {\it A Tour through N=2 strings}, hep-th/9211059;
{\it The N=2 open string}, Nucl. Phys.  {\bf B387}, 263 (1992) [hep-th/9207024].
\bibitem{Mart} E.Martinec, {\it M-theory and N = 2 strings}, hep-th/9710122.
\bibitem{Siegel} W.Siegel {\it The N=4 string is the same as the N=2 string}
Phys. Rev. Lett.  {\bf 69}, 1493 (1992)
[hep-th/9204005]; {\it N=2, N=4 string theory is selfdual N=4 Yang-Mills theory}
Phys. Rev. D {\bf 46} (1992) 3235; ibid D {\bf 47} (1993) 2504; 2512.
\bibitem{Ketov} S.Ketov, {\it Do 
 the critical (2,2) strings know about a supergravity in  2+10 
 dimensions?}, hep-th/9710086; {\it Self-duality and F theory} hep-th/9612171;
 {\it From N = 2 strings to F \& M theory},
 Nucl. Phys. Proc. Suppl.  {\bf 52A}, 335 (1997)
 [hep-th/9606142].
\bibitem{JHEP1} A.Kumar, A.Misra and K.L.Panigrahi {\it Noncommutative
$N=2$ Strings}, JHEP 0102 (2001) 037 [hep-th/0011206].
\bibitem{LM}H.' Liu and J. Michelson, 
{\it *-TREK: The one loop N = 4 noncommutative SYM action},
hep-th/0008205;T.~Mehen and M.~B.~Wise,
{\it Generalized 
*-products, Wilson lines and the solution of the  Seiberg-Witten 
equations}, JHEP {\bf 0012}, 008 (2000) [hep-th/0010204];
M.R. Garousi, Nucl.Phys. B{\bf 579} (2000) 209,
[hep-th/9909214]; Nucl.Phys. B602 (2001) 527-540 [hep-th/0011147].
\bibitem{Ademollo} M. Ademollo  et al., {\it Dual String 
With U(1) Color Symmetry},
Nucl.Phys. B{\bf 111} (1976) 77.
\bibitem{gdf} B. Chen, H. Itoyama, 
T. Matsuo, K. Murakami {\it Correspondence 
between Noncommutative Soliton and Open
String/D-brane System via Gaussian Damping Factor}, hep-th/0010066.
\bibitem{Greensite:1988hm}
J.~Greensite and F.~R.~Klinkhamer,
{\it Superstring Amplitudes And Contact Interactions},
Nucl.\ Phys.\ B {\bf 304}, 108 (1988).
\bibitem{Greensite:1987sm} J.~Greensite and F.~R.~Klinkhamer,
{\it New Interactions For Superstrings},
Nucl.\ Phys.\ B {\bf 281}, 269 (1987);
{\it Contact Interactions In Closed Superstring Field Theory},
Nucl.\ Phys.\ B {\bf 291} (1987) 557.
\end{thebibliography}
\end{document}